\documentclass[a4paper,twoside]{article}
\newcommand{\affil}[1]{$^{\rm #1}$}
\usepackage[authoryear]{natbib}
\bibpunct{(}{)}{;}{a}{}{,}
\usepackage{graphicx,epsfig}
\usepackage[english]{babel}
\date{} 
\title{\large\bf\flushleft Centaurus A at Hard X-rays and Soft Gamma-rays}

\author{\parbox{\textwidth}{\flushleft
\vspace{-0.5cm}
{\it H. Steinle\affil{}}\\
\vspace{0.4cm}
{\small \affil{}\,Max-Planck-Institut f\"ur extraterrestrische Physik, 
                   Postfach 1312, 85741 Garching, Germany}\\
{\small \affil{}\,Email: hcs@mpe.mpg.de}}}
\begin{document}
\twocolumn[
\begin{changemargin}{.8cm}{.5cm}
\begin{minipage}{.9\textwidth}
\vspace{-1cm}
\maketitle
%
%

\small{\bf Abstract:}
     Centaurus~A, at a distance of less than 4 Mpc, is the nearest radio-loud AGN. 
     Its emission is detected from radio to very-high energy gamma-rays. Despite the 
     fact that Cen~A is one of the best studied extragalactic objects the origin of 
     its hard X-ray and soft gamma-ray emission $(100~keV < E < 50~MeV)$ is still  
     uncertain. Observations with high spatial resolution in the adjacent soft X-ray 
     and hard gamma-ray regimes suggest that several distinct components such as 
     a Seyfert-like nucleus, relativistic jets, and even luminous X-ray binaries 
     within Cen~A may contribute to the total emission in the MeV regime that has 
     been detected with low spatial resolution. As the Spectral Energy Distribution 
     of Cen~A has its second maximum around 1 MeV, this energy range plays an 
     important role in modeling the emission of (this) AGN. As there will be no 
     satellite mission in the near future that will cover this energies 
     with higher spatial resolution and better sensitivity, an overview of all 
     existing hard X-ray and soft gamma-ray measurements of Cen~A is presented here 
     defining the present knowledge on Centaurus~A in the MeV energy range. 

\medskip{\bf Keywords:}
galaxies: individual (NGC 5128, Centaurus~A) --- X-rays: galaxies --- 
gamma rays: observations

\medskip
\medskip
\end{minipage}
\end{changemargin}
]
\small

\sloppy

\section{Introduction}
     The elliptical galaxy NCG~5128 is the stellar body of the giant double radio 
     source Centaurus~A (Cen~A) that extends about $10^{\circ}$ on the southern sky. 
     In the inner region it contains a jet with a large inclination 
     ($\sim 70^{\circ}$) to the line-of-sight which is detected in all wavelength 
     bands where the spatial resolution is sufficient. The dust lane, which obscures 
     the nucleus at optical wavelengths, is thought to be the remnant of a recent 
     merger ($10^7$--$10^8$ years ago) of the elliptical galaxy with a smaller spiral 
     galaxy \citep{Thomson1992}. This merger and the subsequent accretion of gas and 
     dust onto the central black hole gives rise to the observed activity of the 
     nucleus (AGN). The supermassive black hole at the center has an estimated mass 
     of $10^7$ to $10^8$ solar masses \citep{Marconi2000,Marconi2001,Silge2005, 
     Cappellari2009}.\\
     Cen~A as an active galaxy is usually classified as a FR I type radio galaxy, 
     as a Seyfert 2 object in the optical \citep{DermerGehrels1995}, and as a 
     'misdirected' BL~Lac type AGN at higher energies \citep{Morganti1992}. It is 
     one of the best examples of a radio-loud AGN viewed from the side of the jet 
     axis \citep{Graham1979,Dufour1979,Jones1996}.\\
     Its proximity of $< 4~Mpc$ \citep{Harris1984,Hui1993,Rejkuba2004} makes Cen~A 
     uniquely observable among such objects, even though its bolometric luminosity 
     is not large by AGN standards. It is the nearest active galaxy and therefore, 
     NGC~5128 is very well studied and frequently observed in all wavelength bands
     \citep{Israel1998}. Its emission is detected from radio to high-energy 
     gamma-rays \citep{Johnson1997,Israel1998,Aharonian2009} making it the only 
     radio galaxy detected in the hard X-ray and soft gamma-ray energy range  
     $(100~keV < E < 50~MeV)$, the MeV regime for short. All other AGN detected in 
     MeV gamma-rays (and identified) are blazars \citep{Collmar2001}.\\
     Historically, Cen~A has exhibited greater than an order of magnitude X-ray 
     intensity variability that has been used to define a low, intermediate, and 
     high luminosity state in X-rays \citep{Bond1996}. In the MeV regime however, 
     Cen~A does not show a similar variability in intensity, but the spectral 
     shape changes between the low and intermediate luminosity states as defined
     in the X-ray regime \citep{Kinzer1995,Steinle1998}. In contrast, in the X-ray 
     energy range below $\sim$100 keV no distinct change of the spectral index 
     (1.7--1.8) is observed \citep{Baity1981,Feigelson1981,Morini1989,Maisack1992,
     Jourdain1993} when the luminosity state changes. It has to be noted, that the 
     detections of variability in soft gamma-rays (and in most older X-ray 
     observations) were made with instruments with spatial resolutions that make it 
     impossible to resolve the 
     components known today. Therefore the sources of the variability can only be 
     determined by indirect methods. \citet{Israel1998} argues that based on 
     correlated variations at hard X-rays and millimeter wavelengths, this emission 
     originates in the nucleus whereas correlated variability in soft X-rays and 
     at 43 GHz point to an origin in the jet.\\
     Observations of gamma-rays in general reveal the most powerful sources and the 
     most violent events in the universe. While at lower energies the observed 
     emission is generally dominated by thermal processes, the gamma-ray emission
     provides us with a view on the non-thermal sources where particles are 
     accelerated to extreme relativistic energies. It is therefore important for 
     our understanding and subsequent modeling of the sources of the emission in AGN 
     to measure the Spectral Energy Distribution (SED) over a wide energy range including 
     gamma-rays and here especially the poorly sampled MeV regime.\\ 
     Because Cen~A is so close and modern instruments can resolve the inner part
     of this AGN in almost all wavelength bands down to parsec level, it has the 
     potential to become a key object for AGN science where models can be uniquely 
     tested. Therefore detailed observations of this galaxy across the 
     electromagnetic spectrum are extremely valuable to improve our understanding 
     of AGNs.\\ 
     The existing MeV data will be for the near future the only measurements of 
     Cen~A in this energy range as no new satellite experiment sensitive in this 
     energy range is beyond the conceptual phase. It therefore seems appropriate to 
     summarize our present knowledge here and to provide references to all important 
     results in this energy range.\\
     As Centaurus~A is such an unique object and so many observations exist, a 
     dedicated web site\footnote{http://www.mpe.mpg.de/Cen-A/} has been set 
     up for this object at the Max Planck Institute for Extraterrestrial Physics 
     (MPE) that provides among other information on Cen~A an up-to-date and complete 
     list of references. This web site, as all the others mentioned in this paper, 
     is continuously updated and is guaranteed to exist for a longer period.\\
     This paper is an extended version of a review presented at a dedicated 
     conference that brought together theoreticians and specialists from all 
     wavelength regimes to discuss 'The Many Faces of Centaurus~A' in June 2009 
     in Sydney, Australia.

\section{The MeV Energy Range}
     Compared to the energy ranges adjacent to the MeV regime, observations
     in the 100 keV to 50 MeV range have much less spatial resolution and the
     instruments used are less sensitive. In soft X-rays and high-energy gamma-rays 
     a spatial resolution of arcseconds is possible today ({\it Chandra}, {\it Fermi}), but the 
     best resolution in the MeV regime achieved so far is about 15' up to 
     several hundred keV by the coded mask telescope {\it SIGMA} \citep{Jourdain1993} 
     and about 4 degrees in the energy range 1--30 MeV by the Compton telescope 
     {\it COMPTEL} onboard the {\it Compton Gamma-Ray Observatory} ({\it CGRO}) 
     \citep{Steinle1998}. 
     The main reason for this unfortunate situation, which prevents a clear 
     identification of the MeV radiation sources, is due to the fact that the 
     interaction probability of gamma-rays with matter in this energy range has a 
     minimum (see Figure 1 in \citet{Gehrels2009}) and reflecting surfaces for 
     telescopes can not be built. Although the interaction probability of MeV 
     gamma-rays with matter is very small, the Earth atmosphere absorbs MeV 
     gamma-rays and thus high altitude balloons or satellites are necessary 
     \citep{Pinkau2009} to carry detectors above the absorbing atmosphere to detect 
     MeV gamma-rays. In addition, the low interaction probability also implies, 
     that detectors have to be massive and heavy, making satellite experiments 
     very costly. As can be seen in Figure 1 in \citet{Diehl2009}, there is a 
     'sensitivity gap' in the MeV region which reflects the fact, that no 
     second-generation experiment with improved sensitivity exists in this energy range. 
     In the adjacent energy regions second- and third-generation instruments with 
     orders of magnitude better sensitivity and resolution have been built. Compton 
     telescopes like {\it CGRO-COMPTEL} or coded mask instruments like {\it SIGMA} 
     and {\it INTEGRAL}
     are still the only proven technique in this energy band although new ideas 
     for MeV telescopes exist \citep{Bloser2009}, but none of them will be 
     realized in the near future.\\
     One important feature of Centaurus A is, that the Spectral Energy Distribution 
     has its second maximum in the MeV energy range (see Figure 3). Thus it is 
     important to measure this region with high significance to derive the spectrum 
     and possible time variability with high accuracy to enable tests of models of 
     the high-energy emission of (this) AGN (e.g. \citet{Ghisellini2005,
     Orellana2009}). In addition, a good spatial resolution to determine the sources 
     of the MeV emission would be ideal, as this is a transition region where the 
     contribution of various sources to the global luminosity changes. In soft X-rays 
     the nucleus, jets, radio lobes and X-ray binaries all contribute to the total 
     luminosity (see the very detailed X-ray images from {\it Chandra} \citep{Kraft2001,
     Kraft2003}), whereas at hard gamma-rays only the inner jet and nucleus and 
     possibly the outer radio lobes are seen (\citet{Hardcastle2009}; 
     {\it Fermi}: \citet{Abdo2009a}; 
     {\it H.E.S.S.}: \citet{Aharonian2009}).

     \section{MeV Observations}
     Since the first satellites were launched in the early 1960-ies, a very large
     number of detectors sensitive to X-rays and gamma-rays were exposed to the 
     radiation above the atmosphere. First hints to an extension of the spectrum 
     of Cen~A into the MeV range were obtained between 1972 and 1981 with various 
     balloon experiments and satellites, when finally the MPI-Balloon Compton 
     experiment detected emission up to 20 MeV from the direction to Cen~A in 
     October 1982 (\citet{Ballmoos1987} and references therein).\\
     In 1990/91 the imaging coded mask telescope {\it SIGMA} on board the {\it GRANAT} satellite 
     detected a point-like source at the position of the nucleus of Cen~A with a 
     spatial resolution of $\sim$15' and a positional uncertainty of 4' and measured 
     a spectrum with a photon index of $\sim$2 in the energy range 40--400 keV 
     \citep{Jourdain1993}. The most accurate spectral information between 50 keV and 
     10 GeV including the MeV regime so far was obtained by the instruments {\it OSSE}, 
     {\it COMPTEL}, and {\it EGRET} onboard the {\it Compton Gamma-Ray Observatory}, that operated 
     more than 9 years from April 1991 to June 2000. During this time, {\it OSSE} observed 
     Cen~A almost 100 times with a time resolution of about one day and a spatial 
     resolution of $4^{\circ}$--$12^{\circ}$ in the energy range 50 keV--10 MeV. 
     {\it COMPTEL} has 
     observed Centaurus~A in the energy range 0.75--30 MeV during 40 {\it CGRO} pointings 
     with a time resolution of 14 days and a spatial resolution of $4^{\circ}$. The 
     highest energy range 50 MeV--10 GeV was covered by {\it EGRET} simultaneous to {\it COMPTEL},
     but Cen~A was only detected below 100 MeV. The {\it EGRET} time resolution was also 14  
     days and the spatial resolution several degrees. All this {\it CGRO} data have been 
     entered into the NASA/IPAC Extragalactic Database (NED)
     \footnote{The NASA/IPAC Extragalactic Database (NED) is at: 
               http://nedwww.ipac.caltech.edu/}
     and are available to the public.\\
     {\it INTEGRAL}, a satellite launched in October 2002 and still operational carries
     two instruments sensitive to MeV gamma-rays: the imager {\it IBIS} and the 
     spectrometer {\it SPI}. Cen~A was observed few times so far for about 6.5 days total 
     but the sensitivities of both instruments do not allow measurements of Cen~A 
     beyond 500--600 keV in a reasonable length of observing time \citep{Collmar2001,
     Rothschild2006,Petry2009}. Due to the relatively low sensitivity, the spatial 
     resolution of nominal 12' of the imager {\it IBIS} could not be exploited for Cen~A 
     in the MeV regime to determine the origin of the gamma-ray emission.

     \subsection{Light Curves}
     X-ray intensity data from Cen~A are available since the late 1960-ies and Cen~A 
     has exhibited variations greater than an order of magnitude in X-rays below 100 
     keV between 1973 and 1983 on time scales of years or less \citep{Bond1996,
     Turner1997,Israel1998}. Later observations of Cen~A have also shown variability 
     on similar time scales \citep{Grandi2003,Rothschild2006}, but no high 
     luminosity state was detected since 1985.
\begin{figure}[h]
\hspace{-0.2cm}
\includegraphics[width=7.8cm, height=6cm, angle=0]{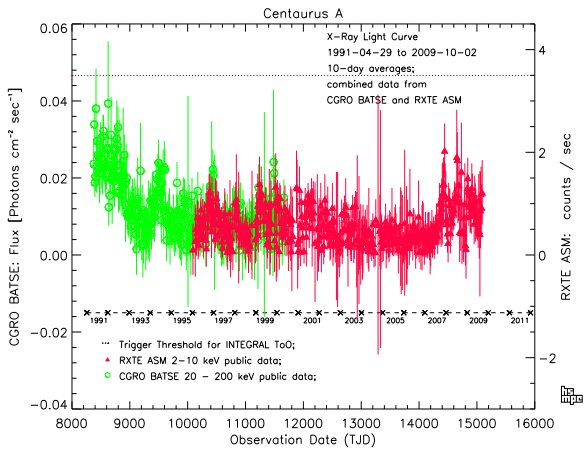}
\begin{center}
\caption{Combined {\it CGRO-BATSE} and {\it RXTE} X-ray light curve spanning 18 years from the 
         launch of {\it CGRO} 1991 to the present (for more details see text). Each data 
         point is a 10-day average.}
\label{fig1}
\end{center}
\end{figure}\\
     \citet{Jourdain1993} distinguish a long-term variability that lasts for
     years and defines the global luminosity state described above, and short 
     intensity variations of the order of days that are superimposed on the 
     long-term component.\\
     X-ray monitoring of Cen~A to define the luminosity state is still pursued today 
     using the all-sky monitors (ASMs) onboard the {\it Rossi X-ray Timing Explorer}
     ({\it RXTE}; 2--10 keV) and on {\it Swift} (15--200 keV).  Fortunately, during
     the time-span from 1995 to 2000, the luminosity of Centaurus~A was monitored 
     simultaneously with the {\it CGRO-BATSE} instrument in the energy range 
     20--200 keV and with {\it RXTE-ASM} in the energy range 1--10 keV so that a 
     combination of both light curves could be achieved. To reduce the 
     statistical noise in the daily data, averages over 10 days are formed for the data
     of both instruments which then show the long-term trends in the luminosity on 
     timescales of months. In Figure 1 the combined data are shown. On the left ordinate 
     the {\it CGRO-BATSE} flux is given whereas on the right ordinate the normalized 
     {\it RXTE-ASM} counts are given. The {\it RXTE-ASM} counts were normalized in such
     a way, that the amplitudes of the variations are similar. No attempt was made to 
     cross-calibrate the measured flux (counts) in the two energy bands.\\
     By comparing the two light curves in the time span of the simultaneous monitoring, 
     it could be verified that the variations observed with the two instruments in 
     different energy bands are almost identical and thus a combined light curve can 
     be established. (A similar normalization is possible with the data simultaneously
     measured at present with the all-sky monitors on {\it RXTE} and {\it Swift}.) It 
     is therefore possible to continue the monitoring and to combine the data sets to 
     create continuous 10-day averages from 1991 to the present (and hopefully also in 
     the future with {\it Swift}.)\\
     Also obvious from Figure 1 is the fact that only at the beginning of these 
     monitoring observations in 1991 an intermediate luminosity state of Cen~A was 
     detected. Since then the source was found to be in a low state. Normalized to 
     the flux at 100 keV (as in \citet{Jourdain1993} and \citet{Bond1996}, which is
     also the middle of the energy ranges of {\it CGRO-BATSE} and {\it Swift}) the 
     intermediate intensity state of Cen~A has a flux of 
     $\sim 6 \times 10^{-5}$ ph cm$^{-2}$ s$^{-1}$ keV$^{-1}$ and the low state has a 
     flux of 
     $\sim 2 \times 10^{-5}$ ph cm$^{-2}$ s$^{-1}$ keV$^{-1}$. The intermediate and low 
     state luminosities in the energy range 40--1200 keV are $\sim 10^{43}$ erg s$^{-1}$  
     and $\sim 4 \times 10^{42}$ erg s$^{-1}$ respectively, assuming a 
     mean single power law with index of 1.8 in this energy range \citep{Bond1996} (and
     ignoring the possible breaks in the spectrum discussed in Section 3.2 below).\\
     Given the low spatial resolution of the instruments used to determine the flux 
     from the direction of Cen~A, it is possible that (superluminal) X-ray binaries 
     in Cen~A contribute significantly to the measured variability at the low energy 
     X-rays that are used to monitor its luminosity state \citep{Steinle2000a,Kraft2001}. 
     Even in the MeV regime a contamination of the luminosity by such objects 
     is possible as e.g. Cyg X-1, a galactic black hole candidate, is detected up to
     MeV energies \citep{McConnell2000}. The time scales of the variations in the 
     monitoring data and that of typical X-ray binaries seem to be similar, 
     supporting the possibility that at least some of the observed variations in 
     the global luminosity may be caused by this class of objects. Although this 
     possible contamination can not be determined for older measurements it is now 
     possible to check the origin of a significant increase in the flux of 
     Centaurus~A as detected by the {\it ASMs} by taking contemporaneous high resolution 
     X-ray images with the satellites {\it Chandra} or {\it Swift}. Therefore, X-ray monitoring 
     is still a tool to define the actual luminosity state of Cen~A.\\
\begin{figure}[h]
\hspace{-0.5cm}
\includegraphics[width=8cm, height=6cm, angle=0]{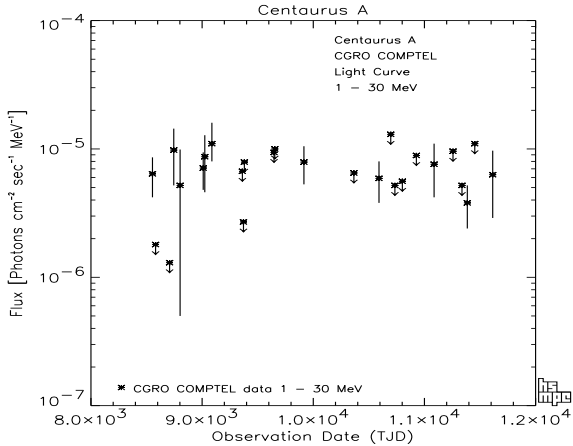}
\begin{center}
\caption{{\it CGRO-COMPTEL} light curve of Cen~A in the 1--30 MeV range 
          spanning 9 years from 1991 to 2000. The data points represent 
          observations of 14 days duration each. Upper limits are two sigma. 
          No significant variability is detected.}
\label{fig2}
\end{center}
\end{figure}
     Probably correlated with the luminosity state as defined by the X-ray monitoring 
     with {\it BATSE} on board {\it CGRO} was the detection of spectral variability in the MeV 
     range by {\it CGRO-COMPTEL} (see section 3.2). This correlation is the main 
     justification to use the X-ray monitoring as a tool to estimate the activity 
     in the MeV energy band as in contrast to the spectral variability observed, no 
     significant luminosity variation was detected with {\it CGRO-COMPTEL} in the MeV 
     range as shown in Figure 2. Also {\it INTEGRAL} did not detect intensity variations 
     in the observed energy range \citep{Bouchet2005,Bouchet2008,Petry2009}. \\
     This non-detection of a luminosity variation in the total energy band 1--30 MeV
     as opposed to the observed variation at 100 keV can be explained by the shift of the 
     spectral breaks that change the shape of the spectrum in such a way that the integrated 
     luminosity in the whole band is not affected.

     \subsection{Spectra}
     Spectra covering the MeV range are sparse. \citet{Jourdain1993} list only
     four measured spectra from the direction towards Cen~A between 1968 and 1991 
     that reach beyond 500 keV. A balloon flight of the Rice University in 1974 
     reached 700 keV and found a single power law with an index of 1.9 
     \citep{Hall1976}. Spectra measured by the {\it HEAO 1} satellite in 1978 in the energy 
     band 2 keV--2.3 MeV required the first broken power law: the index steepens 
     from 1.6 below 140 keV to an index of 2.0 above \citep{Baity1981}. The first 
     spectrum of Cen~A reaching well above several MeV was measured with the MPI 
     Compton telescope in October 1982. In a $4^h.5$ duration balloon flight a 
     power law spectrum in the energy range 0.7--20 MeV was detected with a photon 
     index of $1.4 \pm 0.4$ \citep{Ballmoos1987}. This is a very hard spectrum, and 
     the extrapolated luminosity in the 100 keV region of 
     $10^{-4}$ ph cm$^{-2}$ s$^{-1}$ keV$^{-1}$ would indicate that it was taken during 
     a high luminosity state of Cen~A in the definition of \citet{Bond1996}. However,
     simultaneous measurements with the {\it SMM} satellite that was sensitive in almost 
     exactly the same energy range as the balloon experiment could not verify such a 
     high luminosity state if it would have lasted longer than $\sim$ 8 days 
     \citep{Harris1993}. In addition, in a re-analysis of the balloon data with improved 
     analysis methods and using data on Cen~A from the same balloon flight that were 
     not included in
     the previous analysis it was found that the derived spectral slope remained 
     unchanged, but the intensity of Cen~A was much lower and had a large error so 
     that the emission state could not be derived with certainty (M. Varendorff, 1992, 
     private communication). This data are therefore not included in the final SED 
     in Figure 5.\\ 
     Observations in 1990/91 with the {\it SIGMA} satellite yielded the same power law 
     in the energy band 35--200 keV with photon index $\sim$2 in both observations 
     although in the second observation the intensity of Cen~A had changed by a 
     factor of $\sim$3 \citep{Jourdain1993}. This was the first hint of a constant
     power law index during changes of intensity below $\sim$200 keV.\\ 
     Fortunately Centaurus~A was among the first targets, when {\it CGRO} started regular
     observations \citep{Steinle1998}. The MeV range spectrum
     \footnote{Following \citet{Gehrels1997}, all spectra are shown in the form of
     $\nu F_{\nu}$ plots giving the energy flux per logarithmic interval of 
     frequency to ease the comparison of source luminosities in different wavelength 
     bands --- especially in SEDs.} 
     shown in Figure~3 was measured during Viewing Period 12 (VP 12) in 1991 and it 
     is the only intermediate state spectrum of Cen~A measured simultaneous with 
     all instruments on board {\it CGRO} covering 15 keV--10 GeV. After that, the source 
     declined to its low luminosity state that still persists today.
\begin{figure}[h]
\hspace{-0.5cm}
\includegraphics[width=8cm, height=6cm, angle=0]{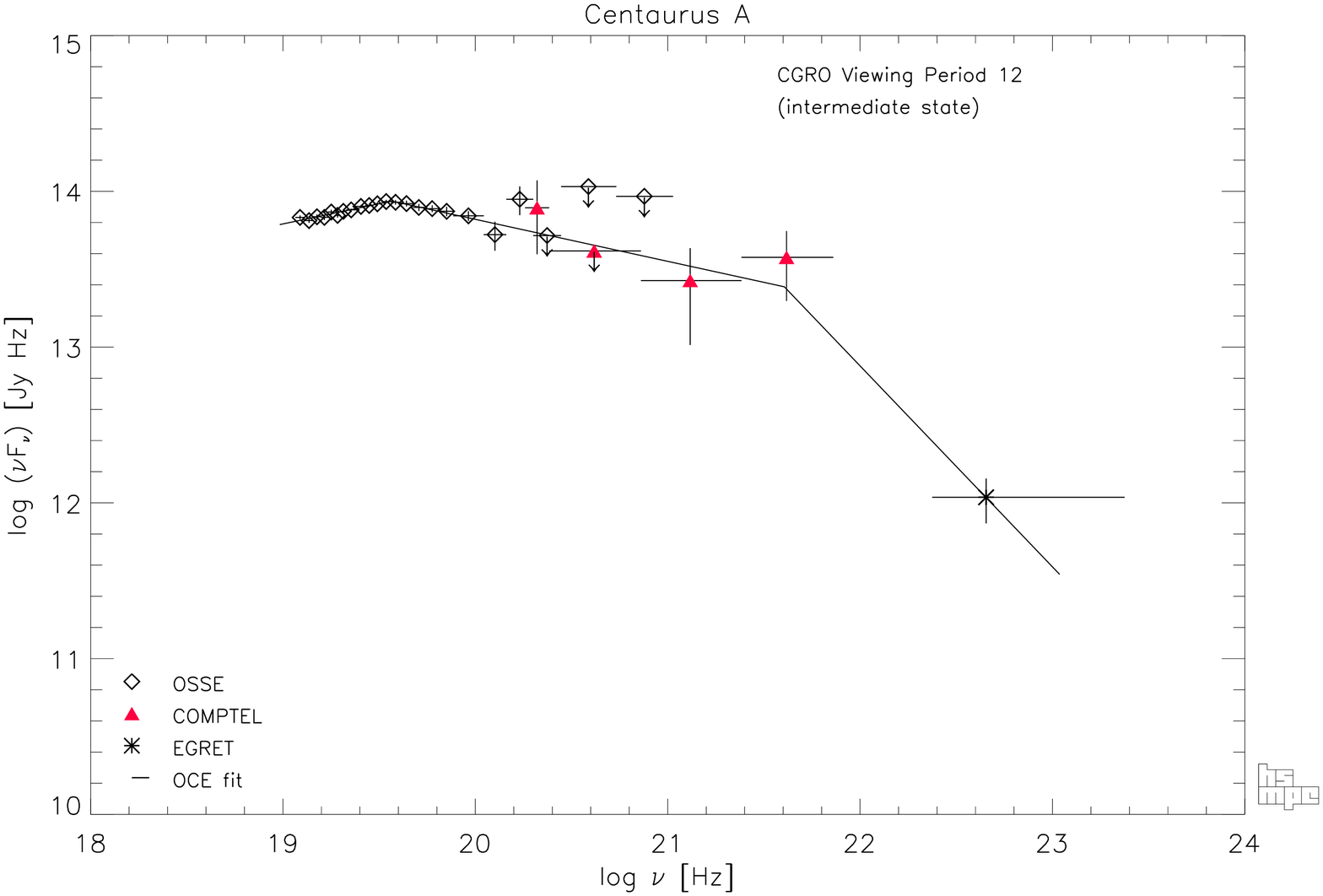}
\begin{center}
\caption{Intermediate luminosity state spectrum of Centaurus~A measured simultaneously 
         with all instruments on board {\it CGRO}. (Viewing Period 12; October 17--31, 
         1991). The data points from {\it OSSE}, {\it COMPTEL}, and {\it EGRET} 
         together with a broken power law fit to the data are shown. See Table 1 for 
         the fit parameter values.}
\label{fig3}
\end{center}
\end{figure}
     Breaks in the power law fit to the data are at 150 keV and 16.7 MeV. As obvious 
     in this $\nu F_{\nu}$ plot, the maximum of the luminosity in this energy range 
     is also at about 150 keV ($3.6 \times 10^{19}$ Hz).\\
     The low state spectrum of Cen~A was measured much more often during the {\it CGRO} 
     mission. It turns out that all those spectra are very similar and can be merged 
     into the average low state spectrum that is shown in Figure 4 together with the 
     MeV range part of the multiwavelength campaign of 1995 that observed Cen~A also 
     in a low luminosity state. In this low state the breaks in the power law are at 
     140 keV and 590 keV and the maximum of the luminosity is shifted to about 600 keV 
     ($1.4 \times 10^{20}$ Hz).\\
     Although the errors in the position (i.e. energy) of break $E_{b_2}$ during the 
     intermediate state are very large (see Table 1) and this state was only observed 
     once with {\it CGRO}, additional information supporting the shift of the break energy 
     towards higher energies during the intermediate state compared to the low state 
     is available. {\it CGRO-EGRET} detects Cen A 
     with good statistics and determines the slope of a power law fit to be 
     $2.85 \pm 0.38$ in its energy range 30--1000 MeV \citep{Nolan1996}. This is 
     consistent within the errors with the spectral index of $\alpha_3 = 3.3 \pm 0.7$ 
     above 16.7 MeV as determined from the whole spectrum in the energy range 30 keV 
     to 400 MeV (Figure 3). Together with the spectral index $\alpha_3 = 2.3 \pm 0.1$ 
     above 150 keV this supports a significant shift in the break energy towards higher 
     energies.\\
     In agreement with most previous measurements reaching higher energies beyond 
     100 keV, the power law slope below $\sim$150 keV in the {\it OSSE} data shows no 
     change of the spectral index ($\alpha \sim$1.7--1.8) below the break (i.e. at 
     lower energies than 150~keV) when the intensity changes (e.g. \citet{Baity1981,
     Feigelson1981,Morini1989,Maisack1992,Jourdain1993,Kinzer1995}. {\it CGRO} for the 
     first time showed that the spectral shape above 150 keV changes between the 
     low and intermediate states \citep{Kinzer1995,Steinle1998}.
\begin{figure}[h]
\hspace{-0.5cm}
\includegraphics[width=8cm, height=6cm, angle=0]{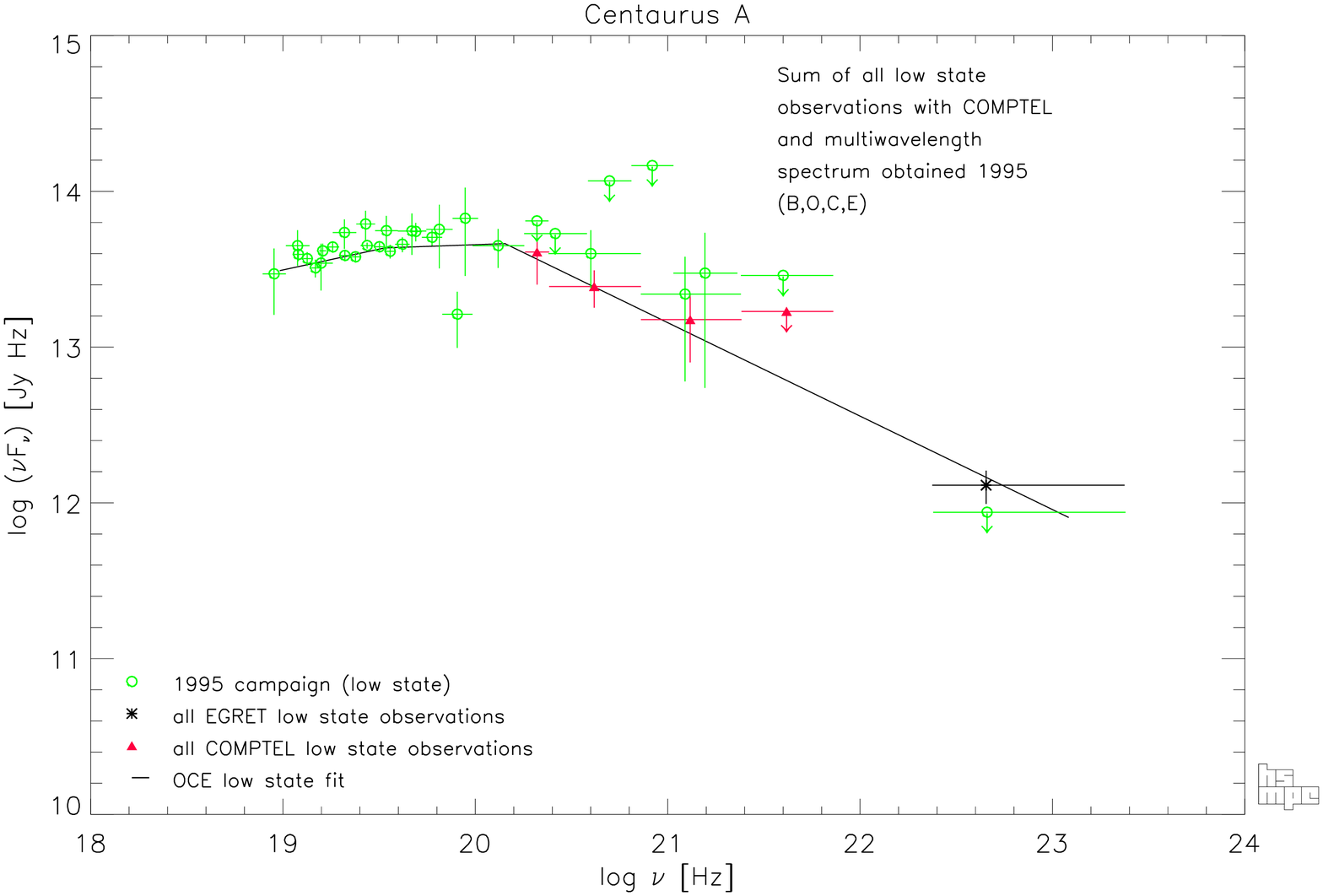}
\begin{center}
\caption{Cen~A low luminosity state spectrum. All {\it CGRO} viewing periods except VP 12, 
         data from the multiwavelength campaign 1995, and the broken power law fit 
         to the data are shown. See Table 1 for the fit parameter values.}
\label{fig4}
\end{center}
\end{figure} 
\begin{table}[h]
\begin{center}
\caption{Spectral Properties of Cen~A }\label{Cen A Spec}
\begin{tabular}{lcc}
\hline Parameter & Intermed. State & Low State\\
\hline  
$E_{b_1}$    & $0.15^{\ +0.03}_{~-0.02}$ MeV & $0.14^{\ +0.03}_{~-0.03}$ MeV  \\
$E_{b_2}$    & $16.7^{\ +27.8}_{~-16.3}$ MeV & $0.59^{\ +0.02}_{~-0.02}$ MeV  \\
$\alpha _1$  & $1.74^{\ +0.05}_{~-0.06}$     & $1.73^{\ +0.05}_{~-0.05}$      \\
$\alpha _2$  & $2.3^{\ +0.1}_{~-0.1}$        & $2.0^{\ +0.1}_{~-0.01}$        \\
$\alpha _3$  & $3.3^{\ +0.7}_{~-0.6}$        & $2.6^{\ +0.8}_{~-0.6} $        \\ 
\hline
\end{tabular}
\end{center}
\end{table}\\
     The low energy low emission state spectra (3--100 (250) keV) measured with 
     {\it INTEGRAL} and {\it RXTE} between 1996 and 2004 \citep{Rothschild2006} confirmed the 
     variability of the emission and the stability of the power law index of 
     $\sim$1.8--2.0. \citet{Petry2009} analyzed all {\it INTEGRAL} observations of the 
     first four years (2003--2007) and produce an average spectrum for the {\it SPI} 
     (25--1000 keV) and for the {\it ISGRI} (25--700 keV) data sets. Both spectra are 
     in excellent agreement and the derived power law index is 1.8--1.9.\\
     To test models of the emission of AGN it is of great importance to measure
     the Spectral Energy Distribution over the widest possible range. Given the fact
     that Cen~A is a very well observed object, its SED can be determined spanning 
     almost 20 decades in energy. Figure 5 shows the final result by combining all 
     measurements contained in the NED, in other publications, and the recently 
     published {\it Fermi} \citep{Abdo2009a} and {\it H.E.S.S.} \citep{Aharonian2009} 
     very-high energy results.
\begin{figure}[h]
\hspace{-0.5cm}
\includegraphics[width=6cm, height=8cm, angle=90]{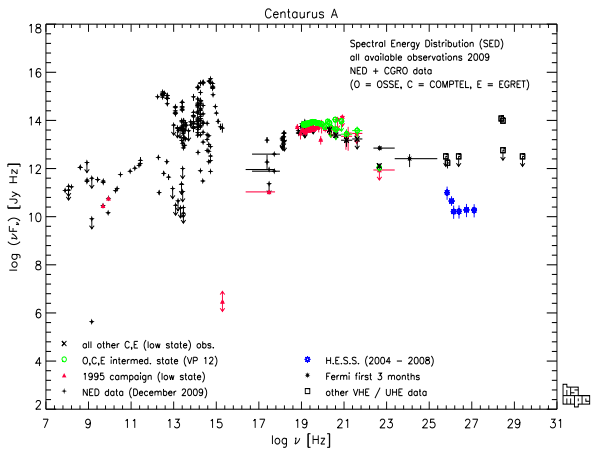}
\begin{center}
\caption{Spectral Energy Distribution (SED) of Cen~A composed from all available 
         data. It has to be noted, that the data in this SED are not contemporaneous. 
         Only the data of the multiwavelength campaign 1995 shown in red filled triangles 
         have been measured simultaneously.}
\label{fig5}
\end{center}
\end{figure}\\
     Indicated in the SED in Figure 5 is the fact, that only for the data in the MeV 
     region discussed in this paper (i.e. $10^{19}$--$10^{23}$ Hz), the luminosity state is 
     known and distinguished. For most of the other data, the luminosity state is 
     not known. As Centaurus~A is variable, the comparison of emission models and 
     the SED requires simultaneous data and specific models for the different
     luminosity states. So far only one simultaneous data set from the 1995
     multiwavelength campaign exists for a low luminosity state \citep{Steinle1999}.

\section{Summary and Outlook}
     The detection of Centaurus~A in gamma-rays up to GeV energies makes this AGN \
     unique, as all other radio-loud AGN detected in high-energy gamma-rays are of 
     the blazar type. As Cen~A is viewed from a large angle with respect to the jet 
     axis, it may well be, that we do not see jet emission from this AGN only 
     (misaligned blazar) but also emission from the nuclear region and that we 
     detect Cen~A only because it is so close.\\
     Observed in the MeV energy range with {\it CGRO} during more than 9 years, and in the 
     following years up to now by {\it INTEGRAL}, Cen~A did not show the large intensity 
     variations in X-rays recorded in the past. Fortunately, at the beginning of the 
     {\it CGRO} observations, Cen~A was in an intermediate luminosity state before its 
     intensity declined to the low luminosity state which lasted for the reminder of 
     the mission and further on until today (October 2009). Only one MeV spectrum 
     has been measured with {\it CGRO} in a luminosity state other than the low 
     state. This intermediate state spectrum differs significantly from the average 
     low state spectrum. During changes in the X-ray luminosity states from 
     intermediate to low, the luminosity in the MeV region stays constant, but
     in contrast to the stability of the spectral shape in the X-ray region, the 
     spectral indices of the broken power law change in the MeV region.\\
     The Cen~A spectra measured with {\it CGRO} close the gap in the MeV energy 
     range that is present in so many other SEDs of AGN due to the lack of 
     high-energy measurements or too low sensitivity in this energy band. This energy
     band however is of specific interest, as almost all emission models of AGN 
     show the second luminosity maximum of the SED to fall in this energy range. 
     When analyzed together with data from other wavelength regimes the 
     {\it CGRO} spectra allowed for the first time to derive a measured 
     continuous Spectral Energy Distribution from the radio to the very-high gamma-rays 
     providing a unique dataset for theoretical modeling of the emission of 
     Centaurus~A over 20 decades in frequency/energy. This SED contains data that
     were collected during very different X-ray luminosity states and it is very 
     important to measure SEDs contemporaneously. Therefore a coordinated 
     multiwavelength campaign was organized in 1995 to measure a simultaneous Cen~A 
     SED in a low luminosity state covering radio to GeV frequencies/energies 
     (see Figure 5).\\
     Many interesting data on Centaurus~A have been collected by the 
     gamma-ray sensitive instruments on various satellites, but still many open 
     questions exist and many important high-energy measurements still have to be 
     made. Among the most interesting observations missing are simultaneous 
     multiwavelength measurements of the SED in a high and intermediate X-ray
     luminosity state, and to determine any possible correlation with the spectral 
     shape in the MeV region, as well as observations with high spatial resolution 
     to resolve jet and nucleus in MeV gamma-rays and to determine the sources of 
     the MeV emission.\\
     To achieve this, second- or third-generation gamma-ray sensitive instruments 
     are needed. In soft X-rays a comparable step was taken going from the 
     {\it Einstein} and the {\it EXOSAT} satellites to the {\it Chandra} and {\it XMM/Newton} 
     observatories. At high-energy gamma-rays this was the step from {\it COS--B} to {\it EGRET}
     and now to {\it Fermi}
     \footnote{A list of high-energy observatories is available from NASAs High 
               Energy Astrophysics Science Archive Research Center (HEASARC):
               http://heasarc.gsfc.nasa.gov/docs/corp/observatories.html}.\\
 
     The outlook is not very encouraging. No new satellite experiment sensitive in 
     the MeV energy range is accepted ({\it ASTRO--H} only reaches up to 600 keV) and only 
     two proposals for large instruments covering the MeV range exit: {\it ACT}, the 
     {\it 'Advanced Compton Telescope'} is an optimized telescope sensitive in the 
     0.2--20 MeV energy band utilizing advances in detector technologies to 
     improve sensitivity by two orders of magnitude, but the spatial resolution will 
     still be only in the order of $1^\circ$ \citep{Boggs2008}.
     {\it GRIPS} (Gamma-ray burst investigation via polarimetry and spectroscopy), mainly 
     aiming towards high spectral resolution, would improve the sensitivity in the 
     200 keV--50 MeV range by a factor of 40 over {\it COMPTEL} but its spatial 
     resolution would still only be 1.0$^\circ$--1.5$^\circ$ \citep{Greiner2009}.\\
     As both projects will at best need many more years to be realized and new ideas
     like the ones listed in \citet{Bloser2009} need much more time, one will have 
     to live for the next decades with what we have --- the data presented in this 
     overview.



\begin{thebibliography}{}
\bibitem[Abdo et al.(2009)]{Abdo2009a}
 Abdo A.A. et al. 2009,
 ApJS, 183, 46
\bibitem[Aharonian et al.(2009)]{Aharonian2009}
 Aharonian F. et al. (H.E.S.S. collaboration) 2009,
 ApJ, 695, L40
\bibitem[Baity et al.(1981)]{Baity1981}
 Baity W.A., et al. 1981, 
 ApJ, 244, 429
\bibitem[von Ballmoos, Diehl, \& Sch\"onfelder(1987)von Ballmoos et al.]
 {Ballmoos1987}
 von Ballmoos P., Diehl R., \& Sch\"onfelder V. 1987,
 ApJ, 312, 134
\bibitem[Bloser et al.(2009)]{Bloser2009}
 Bloser P.F., et al. 2009,
 Astro2010: The Astronomy and Astrophysics Decadal Survey, 
 Technology Development Papers, no. 7
\bibitem[Boggs et al.(2008)]{Boggs2008}
 Boggs S.E., et al. 2008,
 AAS HEAD meeting \#10, \#37.02 
\bibitem[Bond et al.(1996)]{Bond1996}
 Bond I.A., et al. 1996, 
 A\&A, 307, 708
\bibitem[Bouchet et al.(2005)]{Bouchet2005}
 Bouchet L., Roques J.P., Mandrou P., Strong A., Diehl R., Lebrun F.,
 \& Terrier R. 2005,
 ApJ, 635, 1103
\bibitem[Bouchet et al.(2008)]{Bouchet2008}
 Bouchet L., Jourdain E., Roques J.-P., Strong A., Diehl R., 
 Lebrun F., \& Terrier R. 2008,
 ApJ, 679, 1315
\bibitem[Cappellari et al.(2009)]{Cappellari2009}
 Cappellari M., Neumayer N., Reunanen J., van der Werf P. P., de Zeeuw P. T.,
 \& Rix H.-W. 2009,
 MNRAS, 394, 660.     
\bibitem[Collmar(2001)]{Collmar2001}
 Collmar W. 2001, 
 in ESA SP-459, Exploring the gamma-ray universe. 
 Proceedings of the Fourth {\it INTEGRAL} Workshop, 4-8 September 2000, Alicante, Spain,
 ed B. Battrick, A. Gimenez, V. Reglero \& C. Winkler 
 (Noordwijk: ESA Publications Division, ISBN 92-9092-677-5), 241
\bibitem[Dermer \& Gehrels(1995)]{DermerGehrels1995}
 Dermer C.D., \& Gehrels N. 1995, 
 ApJ, 447, 103 and\\ 
 Erratum: 1996, ApJ 456, 412      
\bibitem[Diehl et al.(2009)]{Diehl2009}
 Diehl R., et al. 2009,
 Astro2010: The Astronomy and Astrophysics Decadal Survey, 
 Science White Papers, no. 66, 
 arXiv:0902.2494v1 [astro-ph.GA]
\bibitem[Dufour et al.(1979)]{Dufour1979}
 Dufour R.J., van den Bergh S., Harvel C.A., Martins D.M., Schiffer F.H. III, 
 Talbot R.J.Jr., Talent D.L., \& Wells D.C. 1979,
 AJ, 84, 284
\bibitem[Feigelson(1981)]{Feigelson1981}
 Feigelson E.D., Schreier E.J., Devaille J.P., Giacconi R., Grindlay J.E., 
 \& Lightman A.P. 1981, 
 ApJ, 251, 31
\bibitem[Gehrels(1997)]{Gehrels1997}
 Gehrels N. 1997, 
 Nuovo Cimento B, 112B, 11
\bibitem[Gehrels \& Cannizzo(2009)]{Gehrels2009}
 Gehrels N., \& Cannizzo J.~K. 2009, 
 Exp.Astr., 25, 111
\bibitem[Ghisellini, Tavecchio, \& Chiaberge(2005)Ghisellini et al.]{Ghisellini2005}
 Ghisellini G., Tavecchio F., \& Chiaberge M. 2005,
 A\&A, 432, 401
\bibitem[Graham(1979)]{Graham1979}
 Graham J.A. 1979, 
 ApJ, 232, 60       
\bibitem[Grandi et al.(2003)]{Grandi2003}
 Grandi P., et al. 2003,
 ApJ 593, 160
\bibitem[Greiner et al.(2009)]{Greiner2009}
 Greiner et al. 2009,
 Exp. Astr., 23, 91       
\bibitem[Hall et al.(1976)]{Hall1976}
 Hall R.D., Walraven G.D., Djuth F.T., Haymes R.C., \& Meegan C.A. 1976, 
 ApJ, 210, 631
\bibitem[Hardcastle et al.(2009)]{Hardcastle2009}
 Hardcastle M.J., Cheung C.C., Feain I.J., \& Stawarz L. 2009, 
 MNRAS, 393, 1041      
\bibitem[Harris et al.(1984)]{Harris1984}
 Harris G.L.H., Hesser J.E., Harris H.C., \& Curry P.J. 1984, 
 ApJ, 287, 175            
\bibitem[Harris et al.(1993)]{Harris1993}
 Harris M.J., Share G.H., Leising M.D., \& Grove J.E. 1993, 
 ApJ, 416, 601      
\bibitem[Hui et al.(1993)]{Hui1993}
 Hui X., Ford H. C., Ciardullo R., \& Jacobi G. H. 1993, 
 ApJ, 414, 463
\bibitem[Israel(1998)]{Israel1998}
 Israel F.P. 1998, 
 A\&AR, 8, 237    
\bibitem[Johnson et al.(1997)]{Johnson1997}
 Johnson W.N., Zdziarski A.A., Madejski G.M., Paciesas W.S., Steinle H., 
 \& Lin Y-C. 1997, 
 in: Proceedings of the Fourth Compton Symposium, Williamsburg, VA (USA),
 part I: The Compton Observatory in Review;
 eds. C.D. Dermer M.S. Strickman, and J.D. Kurfess, (New York: AIP),
 AIP Conf.Proc., 410, 283
\bibitem[Jones et al.(1996)]{Jones1996}
 Jones D.L., et al. 1996, 
 ApJ, 466, L63
\bibitem[Jourdain et al.(1993)]{Jourdain1993}
 Jourdain E., et al. 1993, 
 ApJ, 412, 586
\bibitem[Kinzer et al.(1995)]{Kinzer1995}
 Kinzer R.L., et al. 1995, 
 ApJ, 449, 105
\bibitem[Kraft et al.(2001)]{Kraft2001}
 Kraft R.P., Kregenow J.M., Forman W.R., Jones C., \& Murray S.S. 2001
 ApJ, 560, 675
\bibitem[Kraft et al.(2003)]{Kraft2003}
 Kraft R.P., Vázquez S.E., Forman W.R., Jones C., Murray S.S., Hardcastle M.J, 
 Worrall D.M., \& Churazov E. 2003,
 ApJ, 592, 129 
\bibitem[Maisack et al.(1992)]{Maisack1992}
 Maisack M., et al 1992, 
 A\&A, 262, 433
\bibitem[Marconi et al.(2000)]{Marconi2000}
 Marconi A., Schreier E.J., Koekemoer A., Capetti A., Axon D., Macchetto D., 
 \& Caon N. 2000, 
 ApJ, 528, 276      
\bibitem[Marconi et al.(2001)]{Marconi2001}
 Marconi A., Capetti A., Axon D.J., Koekemoer A., Macchetto F.D., 
 \& Schreier E.J. 2001, 
 ApJ, 549, 915 
\bibitem[McConnell et al.(2000)]{McConnell2000}
 McConnell M. L., et al. 2000,
 ApJ, 543, 928     
\bibitem[Morganti et al.(1992)]{Morganti1992}
 Morganti R., Fosbury R.A.E., Hook R.N., Robinson A., \& Tsvetanov Z. 1992, 
 MNRAS, 256, 1p  
\bibitem[Morini, Anselmo, \& Molteni(1989)Morini et al.]{Morini1989}
 Morini M., Anselmo F., \& Molteni D. 1989, 
 ApJ, 347, 750 
\bibitem[Nolan et al.(1996)]{Nolan1996}
 Nolan P.L. et al. 1996, 
 ApJ, 459, 100
\bibitem[Orellana \& Romero(2009)]{Orellana2009}
 Orellana M., \&Romero G.E. (2009)
 AIP Conf. Proc., 1123, 242
\bibitem[Petry et al.(2009)]{Petry2009}
 Petry D., Beckmann V., Halloin H., \& Strong A. 2009,
 arXiv:0909.2802v1 [astro-ph.HE],\\
 accepted for publication in 2009, A\&A  
\bibitem[Pinkau(2009)]{Pinkau2009}
 Pinkau K. 2009, 
 Exp.Astr., 25, 157
\bibitem[Rejkuba(2004)]{Rejkuba2004}
 Rejkuba M. 2004,
 A\&A, 413, 903     
\bibitem[Rothschild et al.(2006)]{Rothschild2006}
 Rothschild R.E., et al. 2006,
 ApJ, 641, 801
\bibitem[Silge et al.(2005)]{Silge2005}
 Silge J.D., Gebhardt K., Bergmann M., \& Richstone D. 2005, 
 AJ, 130, 406     
\bibitem[Steinle et al.(1998)]{Steinle1998}
 Steinle H., et al. 1998, 
 A\&A, 330, 97
\bibitem[Steinle et al.(1999)]{Steinle1999}
 Steinle H., et al. 1999, 
 Adv. Space Res., 23, 911
\bibitem[Steinle, Dennerl, \& Englhauser(2000a)Steinle et al.]{Steinle2000a}
 Steinle H., Dennerl K., \& Englhauser J. 2000,
 A\&A, 357, L57
\bibitem[Thomson(1992)]{Thomson1992}
 Thomson R.C. 1992, 
 MNRAS, 257, 689
\bibitem[Turner et al.(1997)]{Turner1997}
 Turner T.J., George I.M., Mushotzky R.F., \& Nandra K. 1997, 
 ApJ, 475, 118 
\end{thebibliography}
\end{document}